\documentclass[useAMS,usenatbib]{mn2e}
\usepackage[final]{graphics}
\usepackage{amssymb}
\title[On the absence of nova shells]{On the absence of nova shells}
\author[L.~Schmidtobreick, M.~Shara, C.~Tappert, A.~Bayo]
{
L.~Schmidtobreick$^{1}$,
M.~Shara$^{2}$,
C.~Tappert$^{3}$,
A.~Bayo$^{3,4}$,
A.~Ederoclite$^{5}$
\\\\
$^{1}$European Southern Observatory, Casilla 19001, Santiago 19, Chile\\
$^{2}$ Department of Astrophysics, American Museum of Natural History, Central Park West and 79th Street, NY 10024-5192, USA\\
$^{3}$ Instituto de F\'\i sica y Astronom\'\i a, Universidad de Valpara\'\i so, Avda.~Gran Breta\~na 1111, Valpara\'\i so, Chile
\\
$^{4}$ Max-Planck-Institut f\"ur Astronomie, K\"onigstuhl 17, D--69117 Heidelberg, Germany\\
$^{5}$ Centro de Estudios de F\'\i sica del Cosmos de Arag\'on, Plaza San Juan 1, Planta 2, Teruel, E44001, Spain\\
}

\begin{document}

\date{xxxx}

\pagerange{\pageref{firstpage}--\pageref{lastpage}} \pubyear{2014}

\maketitle

\label{firstpage}

\begin{abstract}
We present our wide field H$\alpha$+N[II] observations of 15 
cataclysmic variables to search for remnant nova shells. Such shells have been
found around other cataclysmic variables that were hitherto not known as novae.
Our candidates
were selected as objects in the period regime of high-mass transfer systems
that experience - at least occasionally - low mass transfer rates. 
The fact that we find no indication of a nova shell in any of these systems
allows us to set a lower limit of 13000 years to the recurrence time of
these objects.
\end{abstract}


\section{Introduction}
A nova eruption in a cataclysmic variable (CV) is a thermonuclear explosion 
on the surface of the white-dwarf primary once it has accreted a critical 
mass from its late-type companion. During the nova
eruption, material is ejected into the interstellar medium, 
forming an expanding shell around the CV which can be observed 
once its angular size is sufficiently large to be resolvable from 
the inner binary \citep[see e.g.][]{gill+obrien98-1}.
The total ejected mass is estimated between $10^{-6}$ and $10^{-4} M_\odot$ 
\citep[other references therein]{bode+evans08-1}.

In-between nova eruptions the binary is supposed to appear as a "normal" CV, 
i.e. its behaviour is dominated
by its current mass-transfer rate and the magnetic field strength 
of the white dwarf \citep{vogt89-1}. However, the
model of \cite{sharaetal86-1} predicts that the nova eruption strongly affects the mass-transfer rate $\dot{M}$.
 After an initial phase of enhanced $\dot{M}$ due to the irradiation
of the secondary star by the eruption-heated white dwarf \citep{kovetzetal88-1},
$\dot{M}$ is supposed to decline over the next centuries by several 
orders of magnitude  \citep{livio+shara87-1}, possibly even to the point that
$\dot{M} \sim 0$ due to the secondary star losing contact to its Roche lobe,
a scenario that has been termed "hibernation". 
The recent discovery 
of ancient nova shells around two low-mass transfer systems, 
i.e. Z Cam \citep{sharaetal07-1} and
AT Cnc \citep{sharaetal12-1} provides strong
support for the idea of the nova - dwarf nova cycle originally proposed by
\cite{vogt82-1}. However, even the existence of dwarf novae that have experienced nova outbursts
in the past does not prove the hibernation model. Individual such cases could equally be explained by the fact
that all CVs can explode as a nova once sufficient material has been accumulated on the surface of the white
dwarf. Only a careful population study of novae in comparison to high- and low-mass transfer systems can yield
the answer to this long-debated question.

Most observed old novae do actually show a very high mass transfer rate 
\citep[see e.g.][]{schmidtobreicketal05-1,tappertetal12-1,tappertetal14-1}
This is expected as the recurrence time, i.e. the time between two nova outbursts,
is supposed to be smaller for high mass transfer systems and so they are more likely to be observed during a
nova eruption. Still, this also means that most of the observed old novae are not in 'hibernation' which could
be due to the time-scales and the relatively short time that has passed after the nova eruption (less than 100 years
in most cases).

\begin{table*}
\centering
\begin{minipage}{14.9cm}
\caption{\label{obstab} Summary of the observational details: Object, Filter,
Date \& UT of the first exposure, the number of exposures, the total exposure 
time, the CV subtype \citep{ritter+kolb03-1}, the orbital period, 
and the distance. The references for $P_{\rm orb}$ and $d$ are given below.}
 \begin{tabular}{@{}lccccrclc@{}}
 \hline
 Object & Filter & Date & UT  & \# & $T_{\rm exp}$ [s] & Type & $P_{\rm orb}$\,[d] & d [pc] \\
\hline
2MASS J07465548-0934305 & H$\alpha$ & 2014-03-16 & 01:13 & 7 & 5398 & DN?  & 0.1416 (1) & 180 (1)\\
           & H$\alpha$ & 2013-12-13 & 04:41 & 7 & 5398 & & \\
           & R & 2014-03-16 & 02:55 & 7 & 804 &  &  \\
           & R & 2013-12-13 & 06:23 & 7 & 804 &  &  \\
EF\,Tuc    & H$\alpha$ & 2013-12-12 & 00:40 & 7 & 5389 & DN   & 0.15  (2) & 346 (16)\\
           & R         & 2013-12-12 & 02:23 & 7 & 804  &  &  \\
TU\,Men    & H$\alpha$ & 2014-03-15 & 00:16 & 7 & 5389 & DN   & 0.1172  (3) & 210 (17)\\
           & H$\alpha$ & 2013-12-12 & 02:54 & 7 & 5389 & & \\
           & R & 2014-03-15 & 02:05  & 7 & 804 &  &  \\
           & R & 2014-12-12 & 04:36  & 7 & 804 &  &  \\
2MASS J09400257+2749420 & H$\alpha$ & 2014-03-15 & 02:42 & 7 & 5389 & DN & 0.16352 (4) &\\
           & R & 2014-03-15 & 04:25 & 7 & 804 &  &  \\
SDSS J075939.78+191417.2& H$\alpha$ &   &   & 7 & 5389 & DN & 0.130934 (5) &\\
           & R & 2014-03-17 & 01:49 & 7 & 804 &  &  \\
CTCV J1226-2527 & H$\alpha$ & 2014-03-15 & 05:22 & 6 & 4620 & DN?  & 0.1544 (6) &\\
               &           & 2014-03-16 & 05:31 & 7 & 5389 & & \\
           & R & 2014-03-15 & 06:34 & 7 & 804 &  &  \\
           & R & 2014-03-16 & 07:12 & 7 & 804 &  &  \\
VZ\,Sex    & H$\alpha$ & 2014-03-17 & 02:17 & 9 & 6929 & DN & 0.1487 (7) & 433 (18)\\
           & R & 2014-03-17 & 03:58 & 9 & 1034 &  &  \\
KW2003 105 & H$\alpha$ & 2014-03-15 & 09:21 & 2 & 855 & DN? & 0.1170 (8) & 134 (19)\\
           & H$\alpha$ & 2014-03-16 & 08:09 & 7 & 5389 & & \\
           & R & 2014-03-16 & 09:22 & 7 & 804 &  &  \\
BF\,Ara    & H$\alpha$ & 2014-03-17 & 01:13 & 7 & 5389 & DN & 0.08418 (9) & 758 (19)
\\
           & R & 2014-03-17 & 05:10 & 14 & 1609 &  &  \\
X\,Leo     & H$\alpha$ & 2014-03-16 & 03:22 & 7 & 5389 & DN & 0.1644 (10) & 322 (19)\\
           & R & 2014-03-16 & 05:04 & 7 & 804 &  &  \\
GS\,Pav    & H$\alpha$ & 2014-03-25 & 07:35 & 7 & 5389 & VY\,Scl & 0.155270 (11) & 405 (19)\\
           & R & 2014-03-25 & 09:27 & 7 & 804 &  &  \\
Tau-2      & H$\alpha$ & 2013-12-14 & 03:19 & 7 & 5389 & VY\,Scl & 0.1495 (12) & 787 (19)\\
           & R & 2013-12-14 & 05:01 & 7 & 804 &  &  \\
TT\,Ari    & H$\alpha$ & 2013-12-07 & 01:00 & 7 & 5389 & VY\,Scl & 0.13755 (13)& 335 (20)\\
           & R & 2013-12-07 & 03:05 & 7 & 804 &  &  \\
VZ\,Scl    & H$\alpha$ & 2013-12-13 & 02:04 & 7 & 5389 & VY\,Scl & 0.144622 (14)& 474 (19)\\
           & R & 2013-12-13 & 03:46 & 7 & 804 &  &  \\
1RXS J075330.1+044606 & H$\alpha$ & 2013-12-14 & 05:31 & 7 & 5389 & VY\,Scl & 0.133  (15)\\
           & R & 2013-12-14 & 07:12 & 7 & 804 &  &  \\
 \hline
\end{tabular}\\
{\footnotesize
(1)\cite{pretorius+knigge08-1};
(2)\cite{ritter+kolb03-1};
(3)\cite{mennickent95-1};
(4)\cite{krajci+wils10-1};
(5)\cite{drakeetal10-1};
(6)\cite{augusteijnetal10-1};
(7)\cite{thorstensenetal10-1};
(8)\cite{woudtetal05-1};
(9)\cite{olechetal07-1};
(10)\cite{shafter+harkness86-1};
(11)\cite{grootetal98-1};
(12)\cite{thorstensen+taylor01-1};
(13)\cite{wuetal02-1};
(14)\cite{warner+thackeray75-1};
(15)\cite{sokolovskyetal12-1};
(16)\cite{pretorius+knigge12-1};
(17)\cite{sionetal08-1};
(18)\cite{mennickentetal02-1};
(19)\cite{aketal08-1};
(20)\cite{gaensickeetal99-1}
}
\end{minipage}
\end{table*}

The currently known period distribution of novae shows a significant peak
at 3--4 h \citep{tappertetal13-2} in rough agreement with the theoretical
calculations by \cite{townsley+bildsten05-1}. 
If one considers all types of CVs, this
range between 3 and 4 h is dominated by systems with very high $\dot{M}$
\citep{townsley+gaensicke09-1, schmidtobreick13-1, 2007MNRAS.374.1359R}.
However,
there does exist a small population of dwarf novae in this period range 
\citep{ritter+kolb03-1}.
Also three old novae that have been re-discovered
only recently \citep{schmidtobreicketal05-1, tappertetal12-1}
present optical spectra
that are much more akin to dwarf nova spectra than to nova-like stars. At
least one of them shows also evidence for dwarf-nova like outburst behaviour
\citep[V728 Sco;][]{tappertetal13-1}. 
Could the presence of low $\dot{M}$ CVs in this range
be a consequence of a previous nova eruption as predicted by the hibernation
model? We here present a photometric survey with an H$\alpha$ narrow-band
filter to search for nova shells around CVs in the 3--4 h period range
with low $\dot{M}$ (dwarf novae) or that show occasional low states
(VY Scl stars). The discovery of such nova shells would provide strong
evidence for previous nova eruptions being the cause for the presence of
CVs with relatively low $\dot{M}$ in the 3--4 h period range.

\section{Data and image processing}
We used the Wide Field Imager \citep{baadeetal99-1} at the 
Cassegrain focus of the 2.2-m MPG 
telescope at La Silla, Chile to take deep H$\alpha$+[N\,II] images of
a 30\,arcmin region around the selected CVs. In general, seven 
pointings with small dither offsets to account for the gaps between the chips
were observed. R images were taken to 
subtract the continuum. A summary of the data 
taken for all objects is given in Table \ref{obstab}.

The data reduction was done with THELI \citep{schirmer13-1, erbenetal05-1}, 
including standard bias, overscan, and flat correction. The astrometry 
was calculated with about 50-200 stars for each chip, allowing for a 
{ polynomial of order 3 to fit the}
distortion. For all fields, the internal error was 
$< 0.01^{\prime\prime}$ while the match with the PPMXL catalogue yielded
an accuracy of $\approx 0.31^{\prime\prime}$ which is consistent with the
accuracy of the catalogue positions.

To correct for sky variations between exposures that would show in the mosaic,
a constant sky value was calculated for each exposure and subtracted from each
chip. Then, the average sky value from all exposures of a field were added.
The individual chips of all exposures of one field were combined into the 
mosaic for each filter.
In the end, the mode of data values was used to scale the R-mosaic 
with respect to the narrowband mosaic to be then subtracted as continuum.
{ The resulting mosaic images were inspected by eye to search for 
nova shells.}
\begin{figure*}
\centerline{\rotatebox{-90}{\resizebox{!}{20cm}{\includegraphics{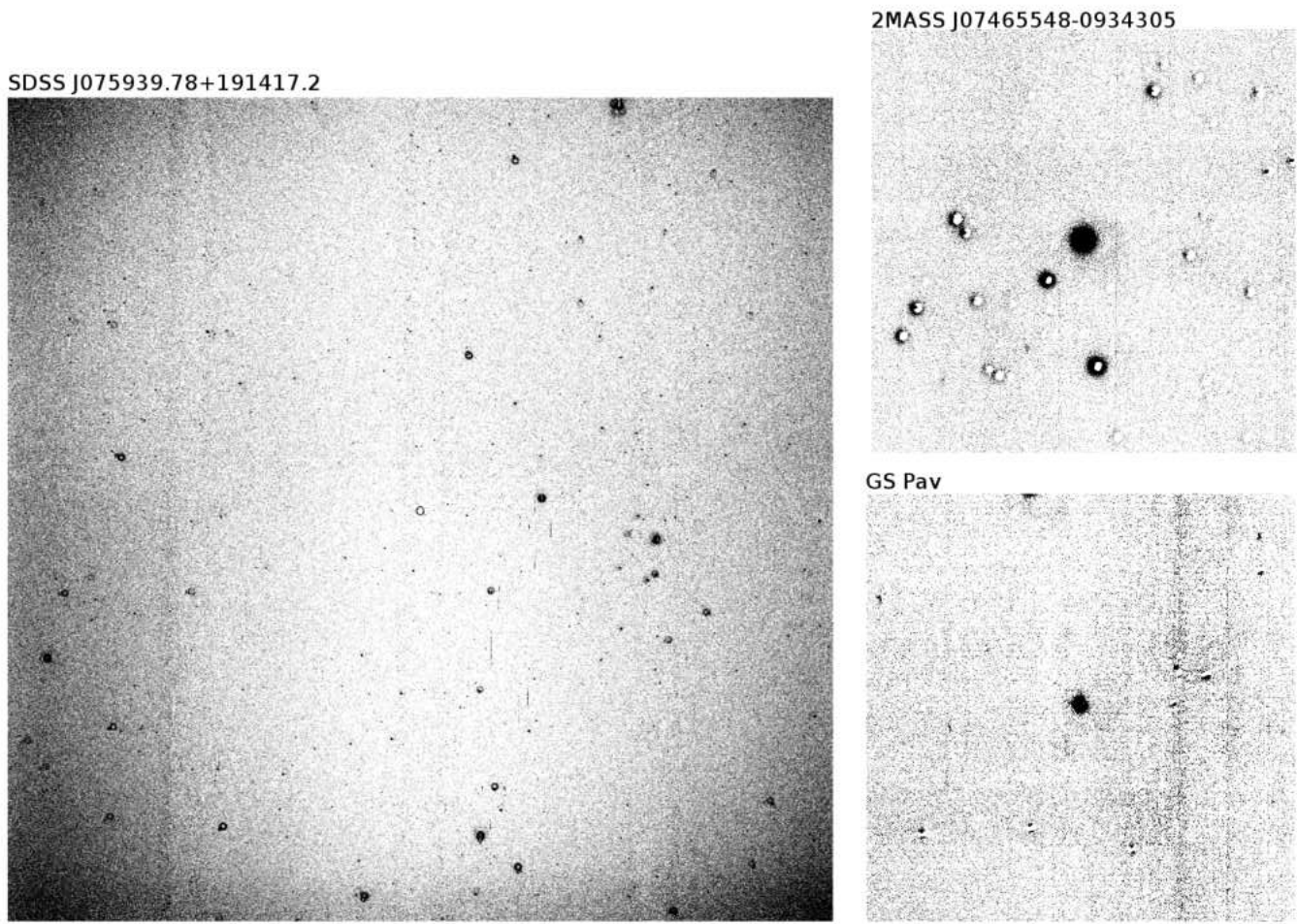}}}}
\caption{\label{mosaic} As examples for the resulting net mosaic images 
(H$\alpha$+[N\,II] minus R) we show a $30'\times 30'$ mosaic of the
field around SDSS\,J075939.78+191417.2 on the left side and 
$2'\times 2'$ zooms onto 
2MASS\,J07465548-0934305 and GS\,Pav on the right side. The contrast has been
set to $\pm$ 2\% of the sky background.
}
\end{figure*}
\section{Results}
Our main result is that no shell is found in any of the analysed fields.
Some examples for the created net (H$\alpha$+[N\,II] minus R)
 mosaic images are given in Figure \ref{mosaic}.
We used ESO's Exposure Time Calculator and compared the results with the
S/N of the average sky to estimate an upper limit for our non-detections. 
With our sensitivity, we 
would have easily ($S/N = 3$ per arcsec$^2$) detected signals as low as 
$\rm 2.5\times 10^{-17} erg\,cm^{-2} s^{-1} arcsec^{-2}$
 and we expect from 
previous detections \citep{sharaetal07-1, sharaetal12-1} that ancient 
nova shells are a factor 5 brighter than this limit. 
\section{Discussion}
From our field of view as well as the resolution and seeing
conditions, we estimate that we should have detected any shell
with a diameter between about 3 arcsec and 30 arcmin.
For objects that are between 200\,pc and 1\,kpc 
we would thus detect { any} shell
with a diameter between 0.02 and 2\,pc. { Note that we might detect
more if the relation between size and distance are advantageous, but
for the size between 0.02 and 2\,pc, we are complete.}
In general, for
the inner, denser parts of shells, velocities between 300 and 500 km/s have been found 
with a few exception of much higher velocities 
\citep[see e.g.][others herein]{sharaetal12-1, downes+duerbeck00-1, gill+obrien98-1}. 
For these kind of shells,
the detection limit due to the size of the detector and the resolution converts
into an age limit of the nova eruption which is between 50 and 2000 years.
We know from
previous observations \citep{sharaetal07-1} that the nova shells
can be visible for these 2000 years, while a nova eruption younger 
than 50 years
should have been observed directly in any of these systems. 

However, not all novae form shells. In fact, 
\cite{cohen85-1} observed 17 old novae, chosen as the brightest and closest 
known, and found 8 shells corresponding to a success rate of about 47\%.
\cite{gill+obrien98-1} observed 17 random novae without known shell and found 
4 shells yielding a success percentage of 24. 
The most complete study was executed 
by \cite{downes+duerbeck00-1} who observed 30 novae and found 13 shells 
which yields again a success rate of 47\%. 
For novae that are further away a better resolution is needed to find 
the shells. So only nearby novae can give an unbiased success percentage.
Alternately, one can use HST imagery as shown by \cite{downes+duerbeck00-1}. 
We thus ignore the low success rate of \cite{gill+obrien98-1} as their data set
is most likely biased with unresolved shells and adopt 
the value of 47\% as the percentage of novae that form a shell.
This value agrees with the complete study of \cite{downes+duerbeck00-1} as well as the one of \cite{cohen85-1} for near novae.

Assuming that all 15 of our observed CVs had a nova eruption during the last 2000 years, 
we should thus have found 7 nova shells on average. The fact that we find none
is thus a strong indication that any nova eruption happened much further in 
the past than we had assumed when starting this project. For an average recurrence time $t_{\rm rec}$, we derive the 
individual probability $p = 0.47 \times \Delta t\times t_{\rm rec}^{-1}$ for a nova to 
explode and produce an observable shell { within the time interval $\Delta t$}. From this, we compute the 
probability that 
none of the novae in our sample shows a shell. We can rule 
out that any nova explosion happened during the last 5000 years with a 3-sigma 
confidence level and push this value to 13000 years with a 1-sigma confidence 
level. This value can thus be regarded as a lower limit for the
average outburst recurrence time.

These findings are consistent with the interval that can be derived from the 
average mass transfer rate of these CVs.
Depending on the mass of the white dwarf as well as its core 
temperature, the accreted mass $M_{\rm ign}$ has to be of the order of 
$10^{-5}M_\odot$ to $10^{-4} M_\odot$  to allow the ignition of the nova explosion \citep{townsley+bildsten05-1}. \cite{townsley+gaensicke09-1} find that the 
average mass transfer rate
for CVs in the 3-4\,h period range that are not SW\,Sex stars is about 
$\rm 5\times 10^{-10} M_\odot/y$. 
The white dwarf in these systems thus needs to 
accrete for at least 20.000 years to reach the critical $M_{ign}$ needed for 
the nova explosion. 

{ In a study of the parameter space for nova outbursts \citep{yaronetal05-1},
also more extreme cases of novae are discussed. For very massive white dwarfs
($M_{\rm WD} \approx 1.4 M_\odot$), they find that only small amounts of 
accreted material $m_{\rm acc}\approx 5\,10^{-8}$ are necessary to ignite
the nova. This seems to be in contradiction to our findings, as this material
should be accreted in the order of 100 years. However, according to their
models, these kind of novae eject a very small amount of material, so might
not be the ones that form shells in the first place. Also, such high white 
dwarf masses are rather rare, so the probability that any such system is among our sample is very low.}

{ The fact that we do not find any nova shells}, also means that, contrary to our initial assumptions, the 
presence of low mass transfer rate systems
in the period regime of the SW\,Sex stars cannot necessarily be explained by 
a modification of the mass transfer rate due to a recent nova explosion. A
larger sample of such stars needs to be analysed to decide with a high
level of confidence whether these
low mass transfer systems actually follow the expected outburst behaviour or not.

%
%

%
\section*{Acknowledgements}
LS thanks the American Museum of Natural History for the hospitality 
during the scientific stay which
made this analysis possible and acknowledges the support through the
ESO DGDF program.
CT acknowledges the funding by FONDECYT Regular grant 1120338.
A. Bayo acknowledges financial support from the Proyecto Fondecyt 
de Iniciaci\'on 11140572.
We gratefully acknowledge the intense use of the SIMBAD data
base, operated at CDS, Strasbourg, France, and of NASA's
Astrophysics Data System Bibliographic Services.
We would like to thank the anonymous referee for valuable comments that helped
to improve the paper.

\bibliographystyle{mn2e}
\bibliography{aamnem99,aabib}

\bsp

\end{document}